\documentclass[pre, reprint, showpacs]{revtex4-1}
\usepackage{amsmath, graphicx, amssymb,color, bm}
\usepackage{ulem}
\newcommand{\jjw}[1]{{\color{black}#1}}

\begin{document}

\title{Effects of passive phospholipid flip-flop and asymmetric external fields on bilayer phase equilibria}

\date{\today}

\author{J.~J.~Williamson}
\email{jjw@fastmail.com}
\affiliation{The Francis Crick Institute, 1 Midland Road, London NW1 1AT, UK}
\author{P.~D.~Olmsted}
\email{pdo7@georgetown.edu}
\affiliation{Department of Physics, Institute for Soft Matter Synthesis and Metrology, Georgetown University, 37th and O Streets, N.W., Washington, D.C. 20057, USA}


\begin{abstract} 
Compositional asymmetry between the leaflets of bilayer membranes modifies their phase behaviour, and is thought to influence other important features such as mechanical properties and protein activity. We address here how phase behaviour is affected by passive phospholipid \textit{flip-flop}, such that the compositional asymmetry is not fixed. We predict transitions from ``pre flip-flop'' behaviour to a restricted set of phase equilibria that can persist in the presence of passive flip-flop. Surprisingly, such states are not necessarily symmetric. We further account for external symmetry-breaking, such as a preferential substrate interaction, and show how this can stabilise strongly asymmetric equilibrium states. Our theory explains several experimental observations of flip-flop mediated changes in phase behaviour, and shows how domain formation and compositional asymmetry can be controlled in concert, by manipulating passive flip-flop rates and applying external fields.

\end{abstract}

\maketitle

\color{red}
\color{black}

\begin{figure*}
\includegraphics[width=18cm]{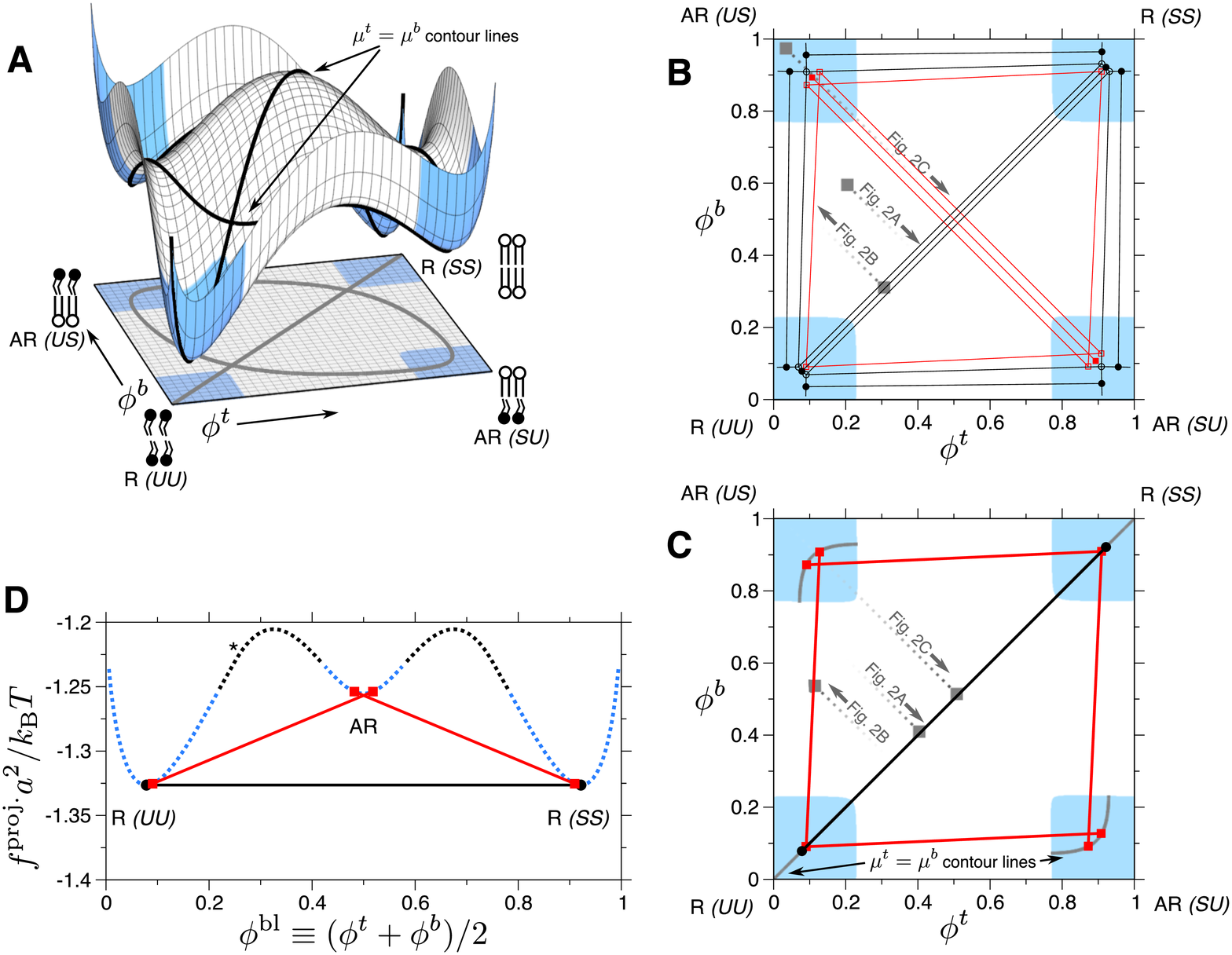}
\caption{\label{phase} A) Landscape of free energy per lipid $f(\phi^t,\phi^b) a^2$ (with $a^2 \sim 0.6\, \textrm{nm}^2$ a typical lipid area) in the space of top and bottom leaflet compositions. 
Compositions in the unstable white region phase separate into registered ``R'' and antiregistered ``AR'' phases lying in the spinodally stable corners (blue).
{The contour {lines} indicate where phospholipid chemical potentials of both leaflets are equal, $\mu^t = \mu^b$}. B) \textit{``Pre flip-flop''} partial phase diagram {for times \jjw{$t < t_\textrm{f-f}$}}, obtained by drawing common tangent planes on $f(\phi^t,\phi^b)$ \cite{Williamson2014}. Equilibrium tie-lines (R-R, R-AR) and triangles (R-R-AR) are black, and metastable {coexistences} (AR-AR,  AR-AR-R) are red. An exhaustive phase diagram including further regions of metastable coexistence may be found in Ref. \cite{Williamson2015}.
C) \textit{``Post flip-flop''} phase diagram at late times. The only allowed states are a single phase lying on the $\mu^t = \mu^b$ contours (cf.\ A) inside the spinodally stable corners, or a tie-line whose endpoints satisfy those criteria. For any other overall leaflet compositions, the bilayer \textit{must} evolve via $\delta \phi^b  = -\delta \phi^t$ to reach an allowed post flip-flop state.
Simulation trajectories for such flip-flop mediated transitions (see Fig.~\ref{trajs}) are shown here as dashed lines that evolve from the initial (a square in B) to the final overall composition (a square in C). 
D) Projected free energy $f^\textrm{proj.}(\phi^\textrm{bl})$ (dotted line) along the $\mu^t = \mu^b$ contour as a function of $\phi^\textrm{bl} \equiv (\phi^t + \phi^b)/2$. {Blue segments} are spinodally stable. At $\phi^\textrm{bl} \approx 0.25$ (asterisk), where the $\mu^t = \mu^b$ contour (see A) splits into diagonal and oval parts, we take the oval path since the diagonal part is spinodally unstable and so irrelevant to phase equilibria. 
The post flip-flop tie-lines from C are indicated as solid lines;\ the \jjw{R-AR states} (red) are doubly degenerate in this projection.
}
\end{figure*}

\section{Introduction}
In model bilayer membranes, phospholipids passively ``flip-flop'' between the leaflets over 
minutes, hours or days \cite{Marquardt2015, Collins2008, Lin2006, Lin2015, Marquardt2017}, in contrast to the much faster translocation of cholesterol or small-molecule additives \cite{Reigada2015}. Membranes of living cells use active, ATP-consuming enzymes to move lipids between leaflets \cite{Fan2008, Turner2005}, but also rely on passive flip-flop via, for example, calcium-activated \textit{scramblase} proteins which allow for more rapid inter-leaflet diffusion than in a pure membrane \cite{Pomorski2004, Brunner2014, Bevers2010}. Synthetic scramblases \cite{Ohmann2017} could allow direct control of passive flip-flop rates in both biological and synthetic membrane systems.

Flip-flop allows the overall compositional asymmetry, \textit{i.e.}, the distribution of phospholipid species between leaflets, to change over time.
Compositional asymmetry is an important parameter in biological function \cite{Zachowski1993, Fadeel2009}, and is important for synthetic membrane applications, changing bilayer rigidity \cite{Elani2015} or influencing the activity of mechanosensitive channels \cite{Senning2014, Perozo2002, Charalambous2012}.
Asymmetry may arise during membrane preparation, or be imposed externally by differing extra-leaflet environments, electric fields or preferential substrate interactions \cite{Wacklin2011, Lin2006, Choucair2007, Stanglmaier2012, Bingham2010}. 

Because of the slow passive phospholipid flip-flop in typical model systems, compositional asymmetry can be prepared and persist over easily-observable timescales \cite{Elani2015, Lin2015, Collins2008}. Therefore, phase-separating mixed model membranes (classically comprising saturated {($S$)} and unsaturated {($U$)} phospholipids plus cholesterol) are amenable to theories that do not include phospholipid flip-flop. {$S$ typically forms liquid ordered ($l_o$) or gel phases (which for our present purpose are interchangable), while $U$ typically forms the liquid disordered ($l_d$) phase.}
``Leaflet-leaflet'' theories (\cite{Wagner2007, Collins2008, Putzel2008, May2009}) assign each leaflet a composition variable which, ignoring phospholipid flip-flop, is separately conserved. 
For typical lateral diffusion coefficient $D\!\sim\!1\,\mu\textrm{m}^2/\textrm{s}$ and flip-flop half-life $\tau_{\scriptstyle \textrm{f-f}}\!\sim\!30\,\textrm{min}$, phase-separated domains could reach a lengthscale $\sqrt{D\tau_{\scriptstyle \textrm{f-f}}}\!\sim\!50\,\mu\textrm{m}$ before flip-flop is important, and such theories indeed explain some features of asymmetric membrane phase behaviour {during this} \textit{``pre flip-flop''} regime \cite{Collins2008, Putzel2008}. 

To our knowledge, the coupling of passive phospholipid flip-flop to phase separation and asymmetry on \textit{``post flip-flop''} timescales has not been studied theoretically. To address this, we extend the leaflet-leaflet description so that flip-flop replaces the separately conserved leaflet compositions by a single conserved \textit{total} composition. We thus predict 
transitions between pre flip-flop and post flip-flop phase equilibria, in the common case where lateral diffusion is much faster than flip-flop. Surprisingly, metastable asymmetric states can persist even in the presence of flip-flop.
We next include a symmetry-breaking external field, which can
stabilise \textit{equilibrium} asymmetric states. Our findings explain several experimental observations: delayed domain formation in asymmetrically-prepared bilayers \cite{Collins2008, Visco2014}, and competing symmetric and asymmetric end-states in phase-separating bilayers on substrates \cite{Lin2006, Goksu2009, Choucair2007}. The framework opens the way to systematically control domain formation and transbilayer asymmetry via the manipulation of passive flip-flop and applied external fields.

\section{Materials and Methods}

In this section we describe the general approach of ``leaflet-leaflet'' phase diagrams, in which the composition of each bilayer leaflet is treated explicitly, as introduced in, \textit{e.g.}, Refs.\ \cite{Wagner2007, May2009, Collins2008, Putzel2008}. We describe our specific implementation of this approach \cite{Williamson2014, Williamson2015, Williamson2015a} and the parameters used in the rest of the paper. In Section \ref{sec:results} we use this approach to develop a theory for the effects of flip-flop and external fields on phase behaviour. 

\subsection{Bilayer free energy} \label{app:modelling}
To derive bilayer free-energy landscapes such as Fig.\ 1A, we use a semi-microscopic lattice model of coupled leaflets \cite{Williamson2014, Williamson2015, Williamson2015a}.
We emphasise, however, that a similar general form of landscape also arises from a fully phenomenological approach \cite{Wagner2007, May2009, Collins2008, Putzel2008}, and most of our findings here follow from very general considerations common to either case. A typical phenomenological free-energy density would take the form (\textit{e.g.}, \cite{Wagner2007})
\begin{align} \label{eqn:phenom}
f_\textrm{phenom.}(\phi^t,\phi^b) = f_l(\phi^t) + f_l (\phi^b) + \Lambda (\phi^t - \phi^b)^2~,
\end{align}
where $f_l$ is some single-leaflet free-energy density, modelled with a random mixing \cite{Wagner2007} or Landau approximation \cite{Putzel2008}, $\phi^{t(b)}$ is the composition in the top (bottom) leaflet, and $\Lambda$ is a leaflet coupling parameter. 

For completeness and to aid in understanding the simulations included here, we recapitulate the main aspects of our alternative, lattice model approach \cite{Williamson2014}. The principal difference from phenomenological approaches on the lines of Eq.\ \ref{eqn:phenom} (Refs.\ \cite{Wagner2007, May2009, Collins2008, Putzel2008}) is an explicit treatment of the local bilayer (or leaflet) microstructure via the thickness of each leaflet. This, for example, allows treatment of hydrophobic mismatch, whose role as an \textit{indirect} inter-leaflet coupling cannot be accounted for in Eq.\ \ref{eqn:phenom} \cite{Williamson2014}.

The lattice Hamiltonian is 
\begin{align}\label{eqn:fun4}
H = &\sum_{<i,j>} ( V_{\hat{\phi}_i^\textit{t} \hat{\phi}_j^\textit{t}}  +  V_{\hat{\phi}_i^\textit{b} \hat{\phi}_j^\textit{b}}) 
+ \sum_{<i,j>} \tfrac{1}{2}\tilde{J} (d_i - d_j)^2 \notag \\
&+  \sum_{i} \tfrac{1}{2}B (\Delta_i)^2 
+  \sum_{i} \tfrac{1}{2}\kappa \left(( \ell^\textit{t}_i - \ell_0^{\textit{t}i})^2 + ( \ell^\textit{b}_i - 
\ell_0^{\textit{b}i})^2                      \right)~,
\end{align}
\noindent where $\hat{\phi}^\textit{t(b)}_i = 1$ if the top (bottom) leaflet at site $i$ contains an $S$ lipid, $\hat{\phi}^\textit{t(b)}_i = 0$ if $U$. The lattice spacing is $a \sim 0.8\,\textrm{nm}$, leading to an area per lipid $a^2 \sim 0.6\,\textrm{nm}^{2}$. The species-dependent ideal hydrophobic lengths are $ \ell_0^{\textit{t(b)}i} = \ell_{S0}$ for an $S$ lipid at the top (bottom) of site $i$, or $\ell_{U0}$ for $U$, and each site is pairwise registered ($SS$ or $UU$) or antiregistered ($SU$ or $US$). In most situations we expect $\ell_{S0}>\ell_{U0}$ due to the greater order of saturated lipid tails. 

The parameter $V \equiv V_{10} - \tfrac{1}{2}(V_{00} + V_{11})$ quantifies intra-leaflet interactions independent of lipid length, such as those between headgroups, \jjw{or the remaining contribution from differences in acyl chain \textit{ordering} for lipids chosen to have the same effective hydrophobic length}.
The ``direct'' coupling $B$ promotes transbilayer symmetry ($SS$ and $UU$ lattice sites) by generating an energy penalty for length (implicitly, tail ordering) mismatch across the midplane, though as discussed below the details of mechanisms underlying such direct coupling are not crucial to our model.
The hydrophobic ``indirect'' coupling $\tilde{J}$ promotes asymmetry ($SU$ and $US$ sites) by penalising mismatch in the bilayer thickness profile. A similar parameter $J \equiv 4\tilde{J}$ appears in the free-energy density derived from Eq.~\ref{eqn:fun4} (see Eqs.~\ref{eqn:fullfann}, \ref{eqn:Xappendix} below) \cite{Williamson2014}. We will primarily use $J$ in what follows.
$\kappa$ penalises variation from species-dependent ideal length, as lipids stretch or compress to reduce the penalties suffered due to $B$ and $J$

$\Delta_0 \equiv \ell_{S0} - \ell_{U0}$ couples to both $B$ and $J$ to control both the indirect and direct inter-leaflet couplings. Hence, varying tail length mismatch alone is approximated by changing $J$, while varying direct coupling alone is approximated by changing $B$. The particular mechanisms leading to direct inter-leaflet coupling need not be specified for our purposes, since $B$ can simply be mapped to an effective value of the inter-leaflet mismatch energy density $\gamma$:\
\begin{align}
\gamma  = \frac{\Delta_0^2 \kappa B}{2 a^2 (\kappa+2B)},
\label{eqn:mismatchenergy}
\end{align}
which has been estimated in the literature \cite{Pantano2011, Risselada2008, May2009, Putzel2011, Blosser2015, Haataja2017} to have values $0.01-1\,k_\textrm{B}T \textrm{nm}^{-2}$. We emphasise that this ``direct'' coupling category describes any source of area-dependent coupling that favours registration, including recently-proposed inter-leaflet coupling via undulations \cite{Haataja2017, Galimzyanov2017}. 

Of course, our ``microscopic'' picture leading to Eq.\ \ref{eqn:fun4} is itself an idealisation, and does not capture the full detail of lipid-level response to thickness mismatch, etc. The aim is to resolve some microstructural effects beyond the reach of Eq.\ \ref{eqn:phenom}, and use estimated values for the elastic parameters (Sec.\ \ref{sec:simulation}) to obtain suitable associated energy scales. For instance, domain line tension estimates from our model lie within the typical experimental range \cite{Williamson2015a}, and a predicted kinetic competition of symmetry and asymmetry within reasonable parameter ranges has been verified in coarse-grained molecular dynamics simulation \cite{Fowler2016}. 

By coarse-graining the lattice model in a mean-field approximation \cite{Williamson2014}, we obtained the following local free-energy density $f$ which determines phase equilibria, as discussed in Sec.\ \ref{sec:results}:
\begin{widetext}
\begin{align} \label{eqn:fullfann}
f(\phi^t,\phi^b) =  
&k_\textrm{B} T \left[ A\ln{A} + (A+1-\phi^t - \phi^b) \ln{(A+1 - \phi^t - \phi^b)} + (\phi^t-A) \ln{(\phi^t-A)} + (\phi^b -A) \ln{(\phi^b-A)} \right] \notag \\
&+\frac{1}{2}\frac{B\kappa\Delta_0^2(\phi^t+\phi^b)}{2B+\kappa}\left(2-\phi^t-\phi^b\right)-\sigma\left(2A+[\phi^t+\phi^b][1-\phi^t-\phi^b]\right)
- 2 V (\phi^t - \phi^b)^2 
-  2 V (\phi^t + \phi^b -1)^2~.
\end{align}
\end{widetext}
The first terms are the entropic contributions of each lattice site type ($SS$, $US$, etc.). The term involving $B$ and $\kappa$ describes the direct coupling energy balanced against lipid stretching. The term proportional to $\sigma$ (see Eq.\ \ref{eqn:Xappendix}) involves the direct and indirect couplings and reflects the cost/benefit of creating registered versus antiregistered lattice sites (\textit{e.g.}, $SS$ versus $SU$). The final terms proportional to $V$ are those for the independent, Ising-like mixing energies in each leaflet. Eq.\ \ref{eqn:fullfann} arises from coarse-graining out (i.e., integrating over) lipid tail-length fluctuations and different microstructural arrangements as constrained by the local compositions $\phi^t,\phi^b$ \cite{Williamson2014}. This leads to a free-energy density that accounts for the simplified bilayer microstructure in our lattice model, and depends on the locally-averaged composition within each leaflet. 

We state again that this level of detail in the free energy, while serving an important role in some applications of the theory \cite{Fowler2016} is not required to generate a landscape with the general shape of Fig.\ \ref{phase}A. Most of the conclusions in the present work are applicable to landscapes generated by either Eq.\ \ref{eqn:fullfann} or Eq.\ \ref{eqn:phenom}.  

We have defined
\begin{align} \label{eqn:ingredients2}
A(\phi^t, \phi^b) \equiv \frac{2\phi^t\phi^b}
{\phi^*+\sqrt{\phi^{*\,2} + 4\phi^t\phi^b(e^{-2\beta\sigma}-1)}}~,
\end{align}
\noindent where
\begin{align}
\phi^* \equiv \phi^t+ \phi^b +  e^{-2\beta\sigma}     (1-\phi^t -\phi^b)~,
\end{align}
\noindent and
\begin{align}\label{eqn:Xappendix}
\sigma &\equiv  - \frac{\Delta_0^2 \kappa^2 (J-B)}{2(2 J + \kappa) (2 B + \kappa)}~.
\end{align}
which is the excess energy needed to create two antiregistered lattice sites relative to two registered ones \cite{Williamson2014}. 

To model an external symmetry-breaking field, we add a linear term
\begin{align}
\label{eqn:fW}
f_\zeta(\phi^t,\phi^b) &\equiv f(\phi^t,\phi^b) - \zeta \phi^t/a^2~,
\end{align}
where $\zeta$ is here the free energy gain per $S$ lipid in the top leaflet.

\subsection{Leaflet-leaflet phase diagrams} \label{sec:leafleaf}

The free-energy landscape leads to a ``leaflet-leaflet'' phase diagram in $(\phi^t,\phi^b)$ space, an approach first introduced in \cite{Wagner2007, Collins2008, Putzel2008, May2009}. Each leaflet has a composition variable describing the fraction of saturated $S$ lipids, an appropriate order parameter for distinguishing liquid disordered from ordered or gel states. The bilayer can split into a coexistence of multiple phases, within which the composition of each \textit{leaflet} is a projection of the phase in $(\phi^t,\phi^b)$ space onto either axis. 

Phase diagrams in $(\phi^t,\phi^b)$ space capture the idea that it is the \textit{bilayer}, comprising coupled leaflets, which phase-separates, and governs the resulting phase behaviour that will be observed in each leaflet. The coupling between leaflets is critical to defining the allowed thermodynamic phases. The common separation of a bilayer with symmetric overall leaflet compositions into two compositionally-symmetric phases is R-R coexistence, where each phase has two leaflets that have the same microstructure; \textit{i.e., a bilayer phase comprising two identical liquid ordered leaflets  coexisting with a bilayer phase comprising two identical liquid disordered leaflets}. For asymmetric overall leaflet compositions on an R-AR tie-line, a registered bilayer phase coexists with an antiregistered one (in which one leaflet is liquid ordered and the other leaflet is liquid disordered). In such cases  the question of whether domains in one leaflet ``induce them in the other'' is more accurately phrased as whether or not the direct inter-leaflet coupling is sufficient that the R-AR tie-lines deviate from vertical or horizontal. If an R-AR tie-line is flat, the composition and `phase' of one leaflet will be uniform between both the R and the AR phases (although the other leaflet changes its composition) \cite{Williamson2015a, Williamson2015}. Similarly, the question of whether one leaflet's composition ``suppresses domain formation in the other'' is the question of whether some given $\phi^t\!\neq\!\phi^b$ takes the bilayer outside any coexistence regions of the leaflet-leaflet phase diagram, so that both leaflets then remain uniform.

This leaflet-leaflet approach has been extensively discussed \cite{Wagner2007, Collins2008, Putzel2008, May2009, Williamson2014}, and is of great utility in organising simulation and experimental observations \cite{Fowler2016, Collins2008}.


\subsection{Parameters and simulation method} \label{sec:simulation}

The lattice model we use to derive the free-energy can readily be simulated (see Fig.\ 2). Varying the inter-leaflet couplings $J$ and $B$, reflecting different balances between hydrophobic mismatch and direct inter-leaflet coupling, can strongly influence kinetic outcomes, such as trapping metastable AR phases \cite{Williamson2015a}, or favouring them in early kinetics or in small domains \cite{Fowler2016}. 

We use a Monte Carlo simulation protocol that resembles spin-exchange dynamics on each leaflet and is as given in \cite{Williamson2015, Williamson2015a}, with the addition of five attempted leaflet-exchange (flip-flop) moves per Monte Carlo Step. This value is low enough to ensure that flip-flop is significantly slower than lateral diffusion, as expected physically. Our simulated flip-flop respects detailed balance, as must be the case for any passive process. We do not consider here the potentially interesting effects of, \textit{e.g.}, spatially-varying flip-flop attempt rates arising from different local compositions, proximity to a scramblase or bilayer defect, etc.

To calculate the theoretical figures (Figs.~1, 3, 4), we employ Eqs.~\ref{eqn:fullfann} and \ref{eqn:fW}, with the microscopic parameters:\ $V = 0.6\,k_\textrm{B}T$, $J = 2\,a^{-2}k_\textrm{B}T$, $B = 0.25\,a^{-2}k_\textrm{B}T$, $\kappa= 3\,a^{-2}k_\textrm{B}T$ and $\Delta_0 = 1\,a \sim 0.8\,\textrm{nm}$, hence $\gamma \approx 0.1\,a^{-2}k_\textrm{B}T$. These parameters are in a range estimated for typical phospholipids, as motivated in Ref.\ \cite{Williamson2014}. The penalty for antiregistration, $\gamma$, is approximately equivalent to the difference in free energy between R and AR phases, see Fig.\ 1D. This has been estimated between $0.01-1\,k_\textrm{B}T \textrm{nm}^{-2}$, with recent theoretical and experimental estimates lying at the lower end of this range \cite{Haataja2017, Blosser2015}.
In all simulations (Fig.~2), we used $V = 0.9\,k_\textrm{B}T$ in an attempt to ensure the same qualitative regime as in the mean-field theory;\ \textit{i.e.}, that $V$ is above the threshold required for demixing in the absence of any other couplings \cite{Williamson2014}. That threshold is $0.5\,k_\textrm{B}T$ in mean-field theory, but $0.88\,k_\textrm{B}T$ in simulation, which incorporates fluctuations \cite{Huang1987}. We set $\kappa= 3\,a^{-2}k_\textrm{B}T$ and a simulation box side length $\mathcal{L} = 100\, a$. 

For Fig.~2A we set $\Delta_0 = 2\,a$, $B = 0.48\,a^{-2}k_\textrm{B}T$ and $J = 0.4\,a^{-2}k_\textrm{B}T$ ($\gamma\approx 0.7\,a^{-2}k_\textrm{B}T$). Relatively large $B$ and small $J$ physically describe species that differ weakly in hydrophobic length but strongly in whichever properties (tail structural order, stiffness, etc.) govern the direct inter-leaflet coupling. This ensures that metastable states do not become trapped by hydrophobic mismatch, allowing us to focus on the transition between the quasi-equilibrium R-R-AR (pre flip-flop) and equilibrium R-R (post flip-flop) states. 

For Fig.~2B we set $\Delta_0 = 1\,a$, $B = 0.24\,a^{-2}k_\textrm{B}T$ ($\gamma\approx 0.1\,a^{-2}k_\textrm{B}T$). A stronger hydrophobic penalty $J = 4\,a^{-2}k_\textrm{B}T$, representing significant tail length mismatch, causes initial phase separation into a metastable AR-AR-R state and further, in this case, inhibits the nucleation of R domains necessary to subsequently equilibrate to R-R \cite{Williamson2015a}. Flip-flop then causes one AR phase to gradually convert to the other, eventually yielding R-AR, \textit{i.e.}, strongly asymmetric overall leaflet compositions. 

For Fig.~2C we again set $\Delta_0 = 2\,a$, $B = 0.48\,a^{-2}k_\textrm{B}T$ and $J = 0.4\,a^{-2}k_\textrm{B}T$. The initial highly asymmetric composition is outside any phase-coexistence region and the bilayer is uniform. However, it does not satisfy equal chemical potential between leaflets, so flip-flop gradually makes the leaflets more symmetric, bringing the overall leaflet compositions onto the R-R central tie-line and yielding domain formation. In fact, the overall leaflet compositions may typically progress through an R-R-AR triangle on the way to the R-R tie-line, transiently exhibiting a combination of R and AR phases.

\section{Results} \label{sec:results}

\subsection{Pre flip-flop phase diagram ($t < \tau_{\scriptstyle \textrm{f-f}}$)}

Here we recapitulate the formalism of a leaflet-leaflet phase diagram \cite{Wagner2007, May2009, Collins2008, Putzel2008, Williamson2014}, whose derivation is discussed in Sec.\ \ref{app:modelling}. {Let $\phi^\textit{t(b)}$ be the local fraction of saturated $S$ phospholipids in the top (bottom) leaflet.} {For a binary mixture, small $\phi^\textit{t(b)}$ is a liquid disordered ($l_d$) phase, and large  $\phi^\textit{t(b)}$} a gel phase. For ternary membranes with cholesterol, or more complex mixtures \cite{Ackerman2015}, {large $\phi^\textit{t(b)}$ typically represents a liquid ordered ($l_o$) phase} \footnote{Such a pseudo-binary mapping is described in \cite{Putzel2008, Williamson2015}, by assuming that the key order parameter is the relative abundance of saturated and unsaturated phospholipids \cite{Putzel2008}, to which 
the rapidly-flip-flopping cholesterol is slaved.}. 

Either a phenomenological Landau theory \cite{May2009, Collins2008, Putzel2008} or a statistical mechanics derivation (Sec.\ \ref{app:modelling}) \cite{Williamson2014} lead to a free-energy density landscape $f(\phi^t,\phi^b)$ as a function of the leaflets' local compositions (Fig.~\ref{phase}A). The four minima {of $f(\phi^t,\phi^b)$} correspond to registered (R) or antiregistered (AR) bilayer phases, {each with  specific compositions} in each leaflet.  {For example, in the registered phase} $UU$ both leaflets are enriched in unsaturated lipids ($l_d$); {while in the antiregistered phase} $SU$ the top leaflet is enriched in saturated lipids ($l_o$ or gel).
The AR minima should normally have higher free energy, as in Fig.~\ref{phase}A, due to an area-dependent inter-leaflet coupling that favours similar compositions in apposing leaflets \cite{Blosser2015, Haataja2017, Fowler2016}.

A bilayer prepared in the white region of Fig.~\ref{phase}A is unstable to  phase separation into two or three coexisting phases within the blue regions. Coexisting phases are defined by common tangent planes touching $f(\phi^t,\phi^b)$ at two or three points \cite{Williamson2014}. 
The partial phase diagram in Fig.~\ref{phase}B contains equilibrium R-R, R-AR and R-R-AR states (a complete phase diagram including further metastable coexistence regions can be found in Figs.\ 13, 14 of \cite{Williamson2015}). The common case of two symmetric phases, within a bilayer prepared with identical overall leaflet compositions, corresponds to an R-R tie-line {connecting $UU$ ($l_d$ in both leaflets) and $SS$ ($l_o$ in both leaflets) phases}. R-R-AR states (where one bilayer phase has mismatching $l_d$ and $l_o$ in each leaflet) appear as {triangles} in the phase diagram. This has been observed and explained in bilayers prepared with asymmetric \jjw{leaflet compositions} \cite{Collins2008}.
\textit{Metastable} coexistences of AR-AR (two asymmetric phases) or AR-AR-R exist if, as in Fig.~\ref{phase}A, the free-energy landscape exhibits AR local minima (see also Appendix \ref{sec:topol}). These metastable phases are favoured by hydrophobic length mismatch between lipid species \cite{Fowler2016}, especially early in the kinetics or for small domains. This can create a  barrier for nucleation to  the equilibrium state \cite{Williamson2015a}.

General discussion of features of leaflet-leaflet phase diagram topologies and how they might be experimentally accessed is given in Appendix\ \ref{sec:topol}.

\subsection{Post flip-flop phase diagram ($t > \tau_{\scriptstyle \textrm{f-f}}$)}

 At late times phospholipid flip-flop allows significant passive inter-leaflet transport. Leaflet compositions can thus vary in the $\delta \phi^b = -\delta \phi^t$ direction, so that only the \textit{``total''} bilayer composition $(\phi^t\!+\!\phi^b)/2\!\equiv\!\phi^\textrm{bl}$ is conserved. This adds a constraint that the leaflet chemical potentials must be equal, $\mu^t = \mu^b$ (where $\mu^{t(b)} \equiv \partial  f / \partial \phi^{t(b)}$)  \cite{Collins2008a}. 
We thus determine the phase diagram of allowed ``post flip-flop'' states (Fig.~\ref{phase}C) graphically, discarding tie-lines from Fig.~\ref{phase}B whose endpoints are not on the $\mu^t = \mu^b$ contour. Projecting Fig.~\ref{phase}A along this contour yields the free energy $f^\textrm{proj.}(\phi^\textrm{bl})$ as a function of the  remaining conserved variable $\phi^{\textrm{bl}}$ (Fig.~\ref{phase}D), which is in fact the most easily accessible composition variable for  a standard experiment in which the same fluorophore is distributed in both leaflets.

The selected coexistences in Fig.~\ref{phase}C comprise a single R-R tie-line and four R-AR tie-lines. There are no three-phase regions (by the Gibbs phase rule the extra constraint allows a triple-\textit{point}, when a tangent can touch all three minima of Fig.~\ref{phase}D). 
A bilayer must adjust its overall composition via flip-flop ($\delta \phi^b = -\delta \phi^t$) to move to one of the allowed tie-lines of Fig.~\ref{phase}C, or to a homogeneous composition that is on the $\mu^t = \mu^b$ contour and inside a spinodally stable corner. Flip-flop does not automatically lead to symmetric leaflets, because the AR minima in the free energy landscape (Fig.~\ref{phase}A) allow $\mu^t = \mu^b$ to be satisfied in AR phases. That is, a region of the bilayer can have $l_o$ and $l_d$ in the apposed leaflets and undergo continuous passive flip-flop without a net exchange of composition between leaflets.
However, in contrast to the equilibrium R-AR states in Fig.~\ref{phase}B, those in Fig.~\ref{phase}C are metastable;\ the R-R tie-line is lowest in free energy and, because of flip-flop, is \textit{accessible}, for any $\phi^\textrm{bl}$ in the phase-separating range (Fig.~\ref{phase}D).


Note that the AR-AR tie-lines from Fig.~\ref{phase}B satisfy the $\mu^t = \mu^b$ contour. However, an AR-AR phase coexistence can spontaneously become a (metastable) single AR phase, by moving its overall composition to one of the tie-line ends. Analogously to domain wall motion in the non-conserved Ising model, the driving force is then not an inter-leaflet chemical potential difference within the bulk phases, but the reduction of interface energy. We exclude the tie-lines for such AR-AR states from Fig.~\ref{phase}C.

\begin{figure}
\includegraphics[width=9cm]{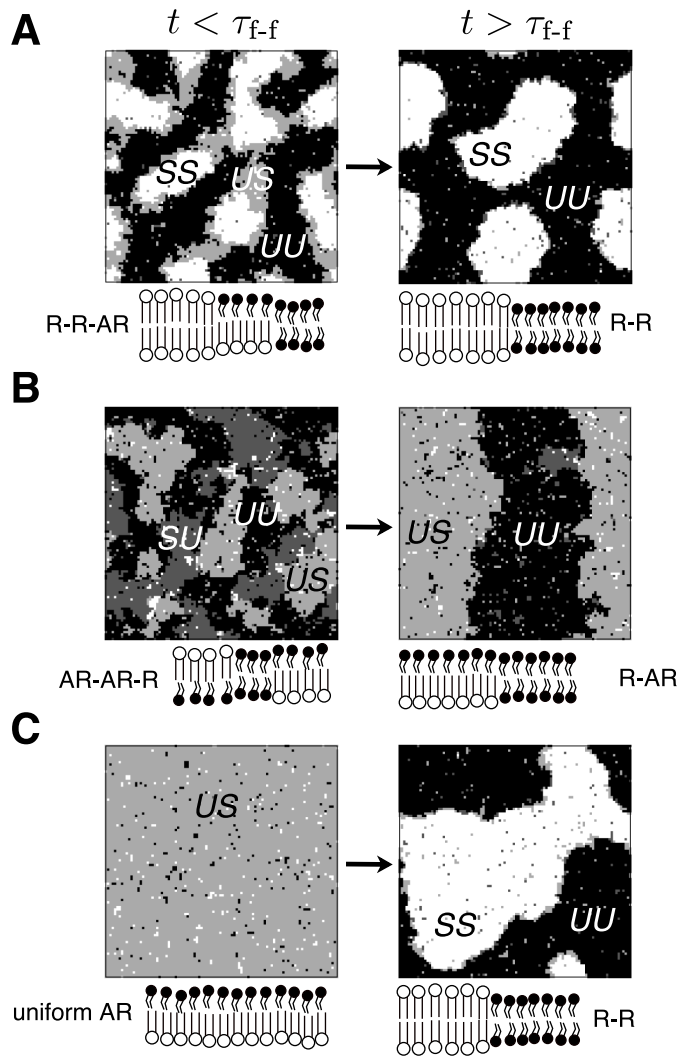}
\caption{\label{trajs} Simulation snapshots illustrating the flip-flop mediated transitions which are labelled in Fig.~\ref{phase}B,C. The initial overall leaflet compositions in A, B, C are $(\phi^{t},\phi^{b})\!=\!(0.2,0.6)$, $(0.3,0.3)$ and $(0.01,0.99)$ respectively, and further details of the model and parameters are given in Section \ref{sec:simulation}. Cartoons beneath each snapshot indicate the coexisting phases present.}
\end{figure}

\subsection{Kinetics of flip-flop mediated transitions}

{For typical situations in which flip-flop is much slower than the diffusion needed to allow lateral phase separation,  we expect transitions between the phase behaviours of Figs.\ \ref{phase}B and \ref{phase}C on timescales longer than the flip-flop time $\tau_{\textrm{f-f}}$.
To illustrate this we perform Monte Carlo simulations of a lattice model of coupled leaflets populated with $S$ and $U$ species \cite{Williamson2015}. The model's free energy has the form shown in Fig.~\ref{phase}A \cite{Williamson2014,SI}, similar to that proposed phenomenologically in \cite{Collins2008, Putzel2008}. Flip-flop moves are attempted at a rate slow enough for domain formation and coarsening to occur before significant flip-flop \cite{Visco2014, Collins2008, Lin2006, Lin2015}. 

Fig.~\ref{trajs}A shows a simulation with {initial overall} composition $(\phi^{t},\phi^{b})\!=\!(0.2,0.6)$. Without flip-flop, the bilayer is in an R-R-AR coexistence region (cf.\ Fig.~\ref{phase}B). Hence, on early timescales we {observe coarsening} of $SS$, $UU$ and $US$ domains. Later, flip-flop eliminates the AR state in favour of symmetric R-R coexistence  at $(\phi^{t},\phi^{b})\!=\!(0.4,0.4)$  (Fig.~\ref{phase}C). The trajectory in $(\phi^{t},\phi^{b})$ {space is} indicated on Fig.~\ref{phase}B,C. 
Hence, R-R-AR {is a} transient, quasi-equilibrium state. Its experimental observation in \cite{Collins2008} {implies that \jjw{flip-flop was slow enough in that case to allow a significant regime of pre flip-flop phase behaviour}.}

Fig.~\ref{trajs}B shows a simulation with {initial overall} composition $(0.3,0.3)$, using  a strong hydrophobic tail length 
mismatch between the lipid species (Section \ref{app:modelling}), which favours metastable AR phases \cite{Fowler2016, Williamson2015a} enough to kinetically trap the metastable AR-AR-R state (Fig.~\ref{trajs}B)}. The coexisting AR-AR phases then decay via flip-flop to a single AR 
phase which coexists with an R phase. In marked contrast to Fig.~\ref{trajs}A, this is a transition to \textit{overall asymmetry}, 
enabled via passive flip-flop and maintained by the inability of the system to nucleate into the equilibrium R-R state. As discussed above, the driving force for this transition is the elimination of AR-AR interfaces, analogously to the non-conserved Ising model \cite{Spirin2001a}.

\begin{figure}
\includegraphics[width=9cm]{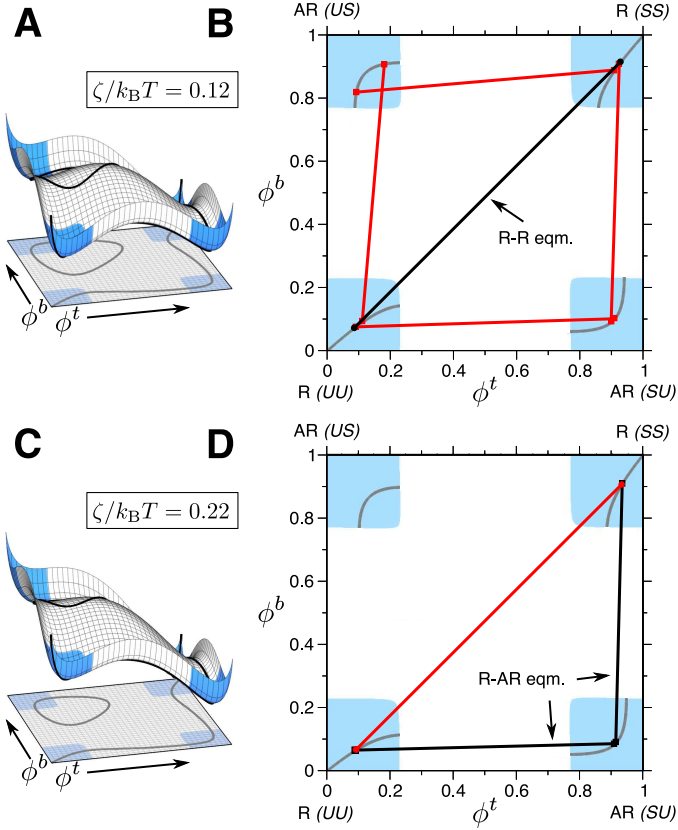}
\caption{\label{external}A,C)\ Free energy landscapes for increasing values of the free energy gain $\zeta$ per $S$ lipid in the top leaflet. The landscape is tilted and the $\mu^t = \mu^b$ contour deformed compared to $\zeta=0$  (Fig.~\ref{phase}A). B) Post flip-flop phase diagram following from A, similarly to Fig.\ \ref{phase}C. D) Phase diagram following from C. The two R-AR tie-lines involving $US$ are no longer allowed, while those involving $SU$ have replaced the R-R tie-line as the equilibrium coexistences.
}
\end{figure}

Fig.~\ref{trajs}C shows a simulation starting with highly asymmetric leaflets, {$(0.01, 0.99)$, within} a single-phase AR region of Fig.~\ref{phase}B such that the bilayer initially remains uniform. Flip-flop induces a transition to the symmetric R-R tie-line of Fig.~\ref{phase}C. This resembles an experiment in \cite{Collins2008}, where an initially asymmetric bilayer displayed domains only after hours, identified as the time required for significant phospholipid flip-flop. A similar phenomenon was attributed to flip-flop in \cite{Visco2014}. Depending on initial composition and phase diagram topology, the path to the two-phase R-R could transiently exhibit R-AR or three-phase R-R-AR states (see Appendix \ref{sec:topol}). 

\subsection{Effect of an external field} 
A variety of external factors can break the up-down symmetry of mixed bilayers, such that one lipid species prefers one leaflet to the other. For example, an electric field transverse to the bilayer may couple to lipids of different charge or headgroup dipole moments \cite{Bingham2010}. Alternatively, the local environments of the leaflets may differ, for example in cells where the plasma membrane leaflets are exposed to cytosol and extracellular fluid. Analogously, in solid-supported bilayers one leaflet is in close contact with the substrate, which can in general be expected to interact differentially with any two lipid species \cite{Wacklin2011, Lin2006, Choucair2007, Stanglmaier2012}. For example, Ref.\ \cite{Lin2006} found that any asymmetric bilayer phases only occurred in one orientation, strongly suggesting a preferential interaction of the substrate with one species.

The simplest effects of such external symmetry-breaking can be modelled \jjw{with an excess free energy per lipid $-\zeta$ for finding one species in one of the leaflets} \footnote{System-specific higher-order effects could be included via nonlinear terms in Eq.~\ref{eqn:modified}.}, \textit{i.e.}
\begin{align}\label{eqn:modified}
f_\zeta (\phi^{t},\phi^{b}) &\equiv f(\phi^{t},\phi^{b}) - \zeta \phi^{t} /a^2 
\end{align}
where $a^2 \sim 0.6\, \textrm{nm}^2$ is a typical lipid area. This external symmetry-breaking field $\zeta$ tilts the free-energy landscape towards high $\phi^t$ (Fig.~\ref{external}A). Since the term is linear in $\phi^{t}$, it does not alter the pre flip-flop phase diagram or affect its stability (white and blue regions).
However, the $\mu^t = \mu^b$ contours and thus the allowed \textit{post} flip-flop states change, since $\mu_t \rightarrow \mu_t-\zeta / a^2$. We expect significant effects when $\zeta$ is comparable to or larger than the free-energy difference between R and AR phases (at $\zeta = 0$), which is $\sim 0.1\,k_\textrm{B}T$ per lipid in Fig.~\ref{phase}D \cite{Williamson2014, SI}. (Recent estimates of this difference are an order of magnitude lower \cite{Haataja2017, Blosser2015}, implying a concomitantly greater sensitivity to a given strength of external field.)

\begin{figure}
\includegraphics[width=9cm]{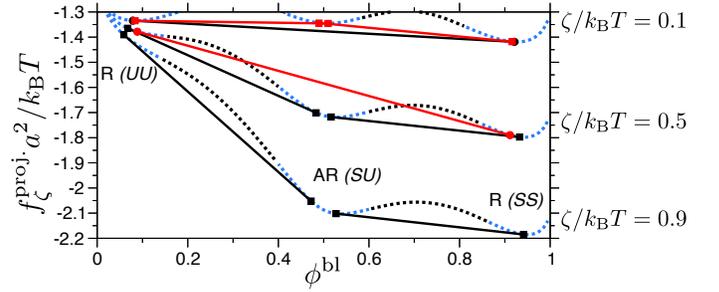}
\caption{\label{external_projected}Projected free energy $f_\zeta^\textrm{proj.}(\phi^\textrm{bl})$ for increasing strength of symmetry-breaking field $\zeta$ (cf.\ Fig.~\ref{phase}B,D). We plot only the branch of the $\mu^t = \mu^b$ contour passing through the $SU$ phase, which is favoured by {$\zeta$} (cf.\ Fig.~\ref{external}). 
{The metastable (red) and equilibrium (black) tie-lines are plotted on each  free energy curve.} A larger  $\zeta$ stabilizes R-AR phase coexistence \jjw{as equilibrium, instead of R-R.}.
}
\end{figure}

Fig.~\ref{external}B shows the post flip-flop phase diagram resulting from Fig.~\ref{external}A for $\zeta/k_BT=0.12$.  The modified $\mu^t = \mu^b$ contour selects different tie-lines compared to the symmetric case ($\zeta=0$, Fig.~\ref{phase}C), but the phase diagram is qualitatively unchanged and R-R remains the equilibrium post flip-flop state. 
For a stronger preference of $S$ for the upper leaflet  ($\zeta/k_BT=0.22$, Fig.~\ref{external}C,D), R-AR tie-lines involving $US$ are absent, while those involving $SU$ are now \textit{equilibrium}, and R-R is only metastable. 

{The trend for increasing $\zeta$ is demonstrated in Fig.~\ref{external_projected}. Similar to Fig.~\ref{phase}D, we plot the projected free energy $f_\zeta^\textrm{proj.}(\phi^\textrm{bl})$ along the $\mu^t = \mu^b$ contour \footnote{Note that Fig.~\ref{external_projected} pertains to the branch of $\mu^t = \mu^b$ going through the $SU$ phase, which is lowest in free energy given positive $\zeta$.}. As $\zeta$ increases, R-AR tie-lines move below R-R to become the lowest free energy states (as in Fig.~\ref{external}D). {For even stronger $\zeta$, fully registered R-R states are completely disallowed. In this limit there is a small region around $\phi^\textrm{bl} \approx 0.5$ where only the homogeneous antiregistered phase $SU$ is allowed, \textit{i.e.}, \jjw{no domains can exist post flip-flop.}}}

In some experiments \cite{Lin2006, Goksu2009, Choucair2007}, liquid-gel bilayers on a substrate converted on a timescale of hours from R-R-AR to either R-R or R-AR but, if deposited directly as R-R or R-AR, remained in the same state. These observations can be explained by our prediction of R-AR and R-R as competing attractors for the \textit{prohibited} R-R-AR state, with a preferential substrate interaction for one species acting to give R-AR coexistence a free energy similar to or lower than R-R (Fig.~\ref{external_projected}).

Appendix \ref{sec:topol} discusses other topologies of the free-energy landscape, aside from those in Figs.~\ref{phase}A,~\ref{external}A,C. For instance, a strong enough direct inter-leaflet coupling favouring registration can remove the AR minima in Fig.~\ref{phase}A \cite{Williamson2014}, so that the $\mu^t = \mu^b$ {locus} {is only a} single diagonal line in the $\zeta = 0$ case. In this case, post flip-flop R-AR states are possible \textit{only} if the external field is sufficiently strong. Conversely, hydrophobic mismatch promotes the existence of AR free-energy minima \cite{Williamson2014}. Therefore, the experimental observations of R-AR competing with R-R post flip-flop \cite{Lin2006, Goksu2009, Choucair2007} might be less likely in systems with weaker hydrophobic mismatch. It is also possible that substrates act to generally increase the effective cost of hydrophobic mismatch, as illustrated in Appendix \ref{sec:coupling}.

\section{Conclusion} \label{sec:discussion}
The framework introduced here models passive phospholipid flip-flop within a simple extension of the ``leaflet-leaflet'' approach to lipid bilayer membrane phase diagrams \cite{Wagner2007, Collins2008, Putzel2008, May2009, Williamson2014}. The analysis and examples provided herein should allow the systematic characterisation of changes in phase behaviour that occur on timescales long enough for passive flip-flop to become important. Specifically, the presence or otherwise of domains in each leaflet, and the total number of bilayer phases, may change over time due to flip-flop, especially in asymmetrically-prepared bilayers or with an external symmetry-breaking field.
Our findings thus suggest the exciting possibility of controlling the transitions to specific post flip-flop states. Flip-flop rates could be controlled by electroporation \cite{Gurtovenko2007, Schwarz1999}, topological defects \cite{Marquardt2017}, or even synthetic scramblase enzymes \cite{Ohmann2017}.

An external field that breaks the bilayer symmetry can be provided by a substrate or, for charged lipid mixtures, an electric field. Charge may have minimal side effects on {miscibility} \cite{Blosser2013}, perhaps {most closely approximating our idealised symmetry-breaking field (Eq.~\ref{eqn:modified})}. The substrates of solid-supported bilayers can act as a symmetry-breaking field in a number of ways:\ a charged substrate will have a preferential interaction with specific charged species and encourage them to be in one leaflet or the other;\ non-charged lipids will naturally have different affinities for a given substrate based on the detailed head-group and surface chemistries, which will similarly encourage compositional segregation across the bilayer.

We have not incorporated hydrodynamics, domain pinning or other anomalous dynamics \cite{Scomparin2009, Garg2007} which, though they cannot affect the free-energy landscape that determines the phase behaviour, may affect the dynamics of reaching the thermodynamically-prescribed state.
This work could be extended to living membranes by combining these passive phospholipid flip-flop effects with \textit{active} lipid recycling via enzymes \cite{Fan2008, Turner2005}.

\section{Author Contributions}
Both authors conceived the research and designed the theoretical model. JJW performed the research and PDO supervised the research. Both authors wrote the paper.

\appendix

\section{Phase diagram topologies}\label{sec:topol}

\begin{figure*}[]
\centering{\includegraphics[width=18cm]{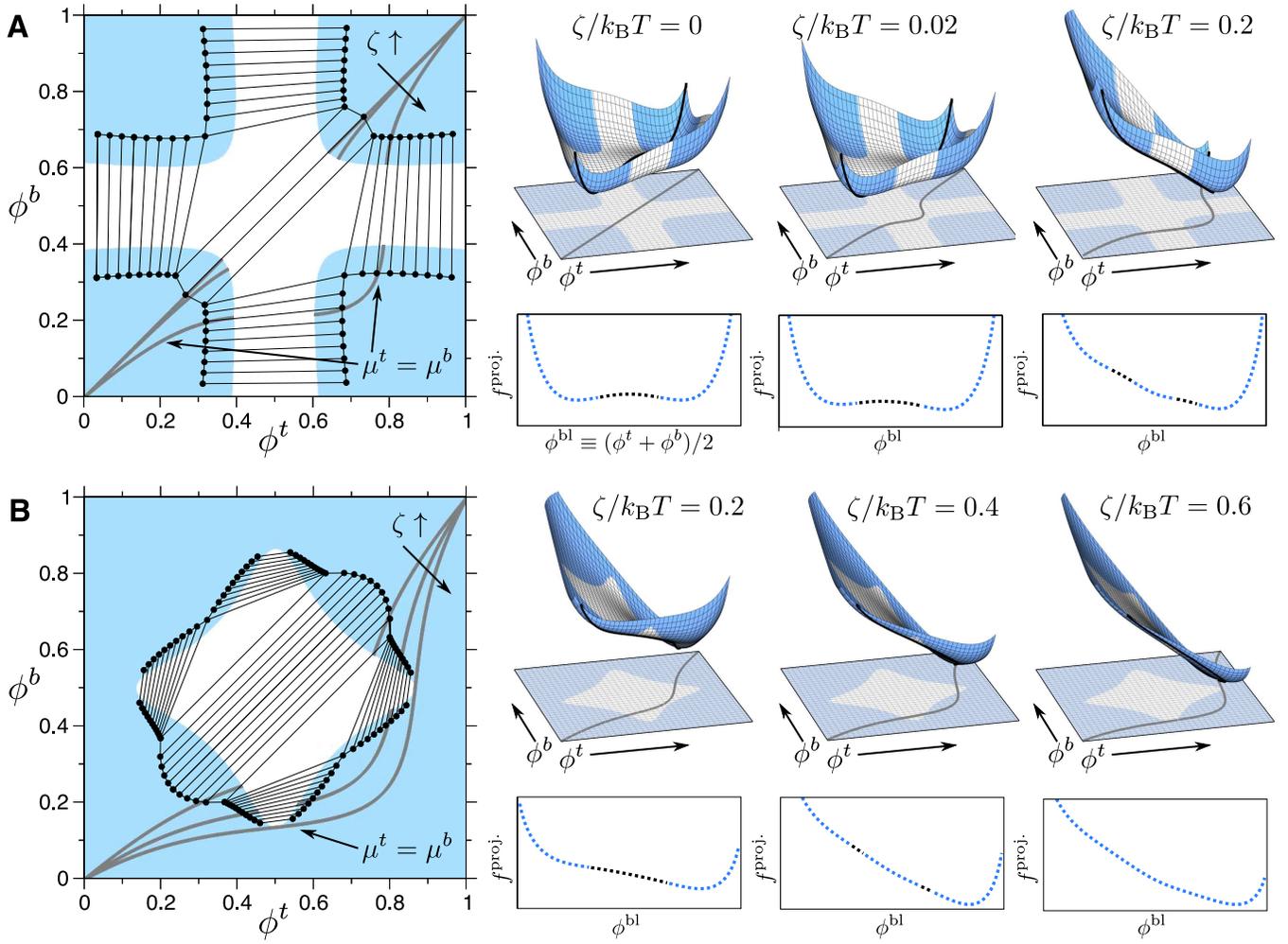}}
\caption{\label{topologies}
Alternative phase diagram topologies (A and B) to that considered in the main text. For each, the pre flip-fop phase diagram is shown, and the $\mu^{t}=\mu^{b}$ contours overlaid (in grey) that pick out the allowed post flip-flop tie-lines (cf.\ Fig.\ 1), for increasing external field $\zeta$ (arrow, cf.\ Fig.\ 3). On the right, for each strength of $\zeta$ a free-energy landscape (cf.\ Fig.\ 3) is shown, along with an illustration of the projected free energy along $\mu^t = \mu^b$ (cf.\ Fig.\ 1D).
A) Parameters as for Fig.\ 1, but with $V = 0.52\,k_\textrm{B}T$ and a higher temperature $T' = 1.19 T$, thus effectively reducing all coupling strengths. There are now no AR free-energy minima and so no metastable AR-AR coexistence. Only the $\phi^t\!=\!\phi^b$ diagonal satisfies $\mu^{t}=\mu^{b}$ for $\zeta = 0$. For $\zeta = 0.02\,k_\textrm{B}T$, $\zeta = 0.2\,k_\textrm{B}T$ (arrow), the $\mu^{t}=\mu^{b}$ contour deforms, for $\zeta = 0.2\,k_\textrm{B}T$ picking out R-AR rather than R-R as allowed post flip-flop tie-lines. 
As is evident in the corresponding projection of $f^\textrm{proj.}$, an R-R tie-line then cannot be drawn satisfying $\mu^t = \mu^b$.
B)\ Now AR minima are absent and, in addition, the four ``arms'' of R-AR coexistence regions become truncated (cf.~\cite{Wagner2007}). Parameters:\ $V = 0.43\,k_\textrm{B}T$, $J = 0.75\,a^{-2}k_\textrm{B}T$, $B = 0.8\,a^{-2}k_\textrm{B}T$, $\Delta_0 = 1\,a$, $\kappa = 3\,a^{-2}k_\textrm{B}T$. The external field is increased (arrow). For $\zeta = 0.2\,k_\textrm{B}T$, the post flip-flop state, if within the phase-separating range, is R-R. For $\zeta = 0.4\,k_\textrm{B}T$, the post flip-flop states comprise two R-AR tie-lines. For $\zeta = 0.6\,k_\textrm{B}T$, the $\mu^t = \mu^b$ contour lies entirely outside the any coexistence region, so only homogeneous post flip-flop state are allowed.
}
\end{figure*}

A variety of topologies are possible for the pre flip-flop leaflet-leaflet phase diagram \cite{Wagner2007, May2009, Putzel2008}. For example, the AR minima in the free energy may be absent if direct inter-leaflet coupling is strong, so that coexistence of two AR phases is impossible, but R-R-AR can still occur because of inflection points in the free energy landscape. Extremely strong direct coupling can even eliminate the R-R-AR three-phase regions, so that only two-phase R-R coexistence is possible;\ this does not appear to apply in the experiments of \cite{Collins2008}, for example.

Hence, a phase diagram topology as in Fig.~1A,B is motivated by the following considerations:\ R minima are lower than AR so that registration is equilibrium;\ R-R-AR coexistence is possible \cite{Collins2008};\ AR minima exist to support AR-AR coexistence \cite{Perlmutter2011, Reigada2015, Fowler2016}. 
We note that AR minima in the free-energy landscape are required in order for the off-diagonal (oval) $\mu^{t}\!=\!\mu^{b}$ contour lines  in Fig.\ 1A to exist. We next discuss two different possible phase diagram topologies, and the resulting consequences for post flip-flop behaviour.

Fig.~\ref{topologies}A shows an alternative phase diagram topology where AR minima are absent, which eliminates metastable AR-AR and AR-AR-R coexistence. 
The off-diagonal $\mu^{t}\!=\!\mu^{b}$ contour lines are now absent. Thus, for $\zeta=0$, only the $\phi^t\!=\!\phi^b$ diagonal satisfies $\mu^{t}\!=\!\mu^{b}$, and the only allowed coexistence \textit{post} flip-flop is R-R. 
An external field $\zeta>0$ deforms the $\mu^{t} = \mu^{b}$ contour and, if this effect is strong enough (here, for $\zeta=2k_BT$), then R-AR tie-lines satisfy $\mu^{t} = \mu^{b}$ while R-R tie-lines do not.
Thus, sufficient $\zeta$ would be required for R-AR to be allowed post flip-flop, whereas in Fig.~1C R-AR was already an allowed, although metastable, post flip-flop state for $\zeta=0$.

As has been previously theorised \cite{Wagner2007}, it is also possible for the four R-AR ``arms'' of the phase diagram to become truncated and narrowed before reaching the edges of the phase diagram. We reproduce this in Fig.~\ref{topologies}B.
In this case, a strong enough $\zeta$ deforms the $\mu^{t} = \mu^{b}$ contour to lie \textit{outside} any phase coexistence region -- no tie-line is allowed post flip-flop. In this case, the external field thus forces the leaflet compositions, \textit{at all values of the total composition} $\phi^\textrm{bl}$, to move outside the binodals, leading to a laterally homogeneous bilayer. 


Our present results imply that the phase diagram topology of a given system can be revealed experimentally by (a)\ preparing an asymmetric bilayer and observing flip-flop mediated changes in phase behaviour;\ (b)\ applying controlled external fields.
In addition, a number of these features can already be inferred in earlier experiments on asymmetric leaflets. For instance, \cite{Collins2008, Visco2014} found that where one leaflet is fully pure (\textit{e.g.}, $\phi^t \sim 0$) no domains appeared in either leaflet, suggesting the composition they used lay outside any coexistence region. This could arise for R-AR two-phase arms that \textit{either} extend to the edges (Fig.~\ref{topologies}A) \textit{or} truncate (Fig.~\ref{topologies}B). To determine which applies, one would need to systematically vary the composition of the other leaflet, to traverse along the $\phi^t \sim 0$ edge of the phase diagram. If no domains form for any composition of the non-pure leaflet, this indicates truncated R-AR arms, whereas R-AR arms that extend to the edges would mean that some range of composition in the non-pure leaflet \textit{would} yield domains. 

Conversely, \cite{Lin2015} finds domain formation even with one leaflet fully pure, implying that in that system the R-AR coexistence regions did not truncate toward the edges of the phase diagram. Egg-sphingomyelin in the non-pure leaflet was found to produce domains in the pure leaflet, based on the exclusion of a certain fluorescent dye.  
Switching to a highly interdigitating milk-sphingomyelin species in the non-pure leaflet caused the domains to also exclude a second type of fluorescent dye, which had not been excluded in the egg-sphingomyelin system. 
This implies that the milk-sphingomyelin increased the direct coupling $B$, which sets the degree to which compositional domains in one leaflet influence the local composition or degree of tail ordering in the other \cite{Williamson2015}, \textit{i.e.}, the degree to which R-AR tie-lines are tilted from horizontal or vertical.
In turn, this suggests important sensitivity of the effective value of $B$ to easily tuneable molecular properties. Similarly, the strong dependence of line tension on hydrophobic mismatch \cite{Garcia2007} implies significant variations in the effective $J$ can be readily achieved. 

\section{Coupling between a solid substrate and bilayer thickness mismatch} \label{sec:coupling}

\begin{figure}[]
\centering{\includegraphics[width=8.5cm]{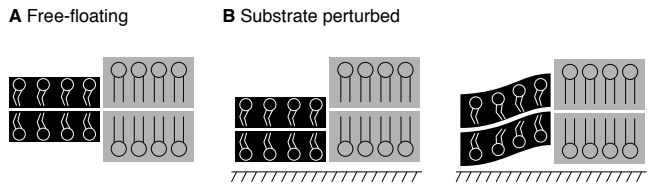}}
\caption{\label{substrate} 
Schematic illustration of a possible mechanism by which hydrophobic mismatches incur a higher penalty in bilayers on a substrate versus free-floating. In A, the equilibrium state, by symmetry, can be presumed to have the thickness mismatch distributed evenly as shown. If a substrate encourages one side of the bilayer to lie flat against it, as in B, this may require additional deformation relative to the state shown in A. 
}
\end{figure}

In addition to a symmetry-breaking effect represented in the parameter $\zeta$, we suggest that hard substrates can increase the penalty for hydrophobic mismatch, increasing the effective value of $J$ and thus promoting asymmetric states in supported bilayers \cite{Lin2006, Goksu2009, Choucair2007}. This is illustrated in Fig.~\ref{substrate}. The bilayer will tend to adhere to the substrate, and at a thickness-mismatch boundary this may lead to a situation resembling one of those in Fig.~\ref{substrate}B. Either case is likely to incur a greater energy cost for the thickness mismatch than the free-floating bilayer shown in Fig.~\ref{substrate}A.

Clearly, the details of any such mechanism would warrant study in their own right. As discussed in \cite{Blosser2015}, the relation between the geometry of thickness mismatch in supported bilayers and that in free-floating ones is not fully understood and is a matter of active research \cite{Nielsen2013, Chen2007}.

Hence, a substrate could encourage R-AR states in two cooperative ways:\ by inducing a species-preferential interaction $\zeta$, and by effectively increasing the hydrophobic mismatch cost that would tend to favour AR free-energy minima in general. It is plausible that this second effect helped contribute to the R-AR final states observed for substrate-supported bilayers in Refs.\ \cite{Lin2006, Goksu2009, Choucair2007}.

\acknowledgments
We acknowledge discussions with MD Collins, A Lamberg, S Redner, SL Veatch, HP Wacklin and members of the EPSRC (UK) CAPITALS programme grant, and thank MC Blosser, SL Keller and ML Longo for input on an early version of the manuscript. Support from the Ives endowment (PDO) and Georgetown University (JJW) is gratefully acknowledged.

\bibliography{bibliography}
\end{document}